\documentclass[%
reprint,
amsmath,amssymb,
 ]{revtex4-1}

\usepackage{graphicx}
\usepackage{upgreek}
\usepackage{xcolor}
\setcitestyle{super}

\begin{document}

\title{\Large\bfseries\noindent\sloppy \textsf{Dual-Comb Real-Time Molecular Fingerprint Imaging }}

\author{Thibault~Voumard$^{1}$}
\thanks{These authors contributed equally}
\author{Thibault~Wildi$^{1}$}
\thanks{These authors contributed equally}
\author{Victor~Brasch$^{2}$}
\author{Raúl~Gutiérrez~Álvarez$^{3}$}
\author{\\Germán~Vergara~Ogando$^{3}$}
\author{Tobias~Herr$^{1}$}
 \email{tobias.herr@cfel.de}
\affiliation{
$^1$Center for Free-Electron Laser Science, Deutsches Elektronen-Synchrotron, Notkestr. 85, 22607 Hamburg, Germany\\
$^2$Time and Frequency, Centre Suisse d'Electronique et de Microtechnique (CSEM), Jaquet-Droz 1, 2000~Neuchâtel, Switzerland\\
$^3$New Infrared Technologies (NIT), Calle Vidrieros 30, Boadilla del Monte, 28660 Madrid, Spain}

\maketitle

% Abstract/Intro
\textbf{Hyperspectral imaging provides spatially resolved spectral information. Utilising dual frequency combs as active illumination sources, hyperspectral imaging with ultra-high spectral resolution can be implemented in a scan-free manner when a detector array is used for heterodyne detection. However, relying on low-noise detector arrays, this approach is currently limited to the near-infrared regime.
Here, we show that dual-comb hyperspectral imaging can be performed with an uncooled near-to-mid-infrared detector by exploiting the detector array's high frame-rate and the combs' high-mutual coherence.
The system simultaneously acquires hyperspectral data in 30~spectral channels across 16'384 pixel, from which molecule-specific gas concentration images can be derived. Artificial intelligence enables rapid data reduction and real-time image reconstruction. Owing to the detector array's sensitivity from 1~$\upmu$m to 5~$\upmu$m wavelength, this demonstration lays the foundation for versatile imaging of molecular fingerprint signatures across the infrared wavelength-regime in real-time.
}

\section{Introduction}
Hyper-spectral imaging extends traditional imaging approaches by providing detailed spectral information for each pixel of an image \cite{bannon2009}. As a general method it has been employed with impressive success across the scientific disciplines, including Earth remote sensing \cite{goetz1985} and medical sciences \cite{lu2014}. 
Hyperspectral imaging instruments often rely on conventional visible or near-infrared photo-cameras in conjunction with dispersive elements or filters. Achieving fast image acquisition as well as high spectral resolution in a large number of pixel remains challenging. Indeed, resolving the narrow optical absorption features of gases requires precision spectroscopic techniques, which typically do not offer spatial resolution. Having the capability to rapidly image and identify characteristic molecular fingerprints of gas molecules would open new opportunities for medical imaging diagnostics, environmental monitoring or industrial applications, including leak detection, process optimisation or identification of hazardous substances. %In these domains, on-field practices would benefit from a portable implementation.

In order to permit rapid and reliable hyperspectral imaging of gas molecules it is therefore desirable to implement precision spectroscopic techniques that combine multiplexed spatial and spectral acquisition. One particularly attractive approach is performing dual-frequency comb spectroscopy with an imaging detector array \cite{martin-mateos2020}, which enables pixel-wise parallel spatial multiplexing and hence rapid acquisition of hyperspectral data without any moving mechanical parts.

In dual-frequency comb spectroscopy\cite{coddington2016, picque2019, schiller2002, keilmann2004, schliesser2005}, two optical frequency combs (1 and 2) are used. Each represents a well defined set of lasers lines spaced by their respective repetition rate $f_\mathrm{rep}^{(1)}$ and $f_\mathrm{rep}^{(2)}=f_\mathrm{rep}^{(1)} + \Delta f_\mathrm{rep}$ ($\Delta f_\mathrm{rep} \ll f_\mathrm{rep}^{(1,2)}$) with a relative frequency offset $f_\mathrm{c} \ll f_\mathrm{rep}^{(1,2)}$ between both combs.
Simultaneous photo-detection of the combined combs with a single detector results in a multi-heterodyne signal comprised of periodic interferograms, each with a duration of $\Delta f_\mathrm{rep}^{-1}$. Fourier-transforming (at least one of) the interferograms yields the multi-heterodyne spectrum comprising beatnotes at frequencies $f_\mathrm{c} + n\cdot\Delta f_\mathrm{rep}$ ($n=0, \pm1, \pm2, ...$). Effectively, the optical spectrum is compressed by a factor of $(f_\mathrm{rep}^{(1)} + f_\mathrm{rep}^{(2)})/(2\Delta f_\mathrm{rep})$ and down-converted from the optical domain to multi-heterodyne frequencies around $f_\mathrm{c}$. Direct dual-comb hyperspectral imaging is achieved when a dual-comb light source illuminates a sample and then is imaged on an 2-dimensional detector array where \textit{each} pixel performs a multi-heterodyne detection. In this way, recent work demonstrated dual-comb hyperspectral imaging with a frame rate of 1~Hz and 5 spectral sampling points using a low-noise indium-gallium-arsenide (InGaAs) near-infrared detector array with a frame rate of 24~Hz \cite{martin-mateos2020}. In contrast to mid-infrared capable detection arrays, InGaAs-based arrays offer low-noise detection, however, are limited to near-infrared wavelength and cover only a small portion of the infrared molecular fingerprint regime.
\\
\\
Here we show that an uncooled, near-to-mid-infrared lead-selenide (PbSe) photo-detector array, sensitive across the entire 1~$\upmu$m to 5~$\upmu$m wavelength range, can be used to perform dual-comb hyperspectral imaging.
Although the array exhibits significant low-frequency Flicker noise, its exceptionally high frame rate of up to 4~kHz permits heterodyne detection above the Flicker noise dominated band. The high mutual coherence of the employed optical combs allows dense spacing (down to the Hz-level) of the multi-heterodyne beatnotes and, in addition, can be leveraged to lower the contribution of thermal detector noise to the final signal. Hyperspectral images with approximately 30 high-resolution spectral channels are recorded with a hyperspectral frame rate of up to 10~Hz. Importantly, the array's near-to-mid-infrared sensitivity allows for direct operation in the molecular fingerprint region. Moreover, a second major challenge arising from the massively parallel data acquisition is addressed in this work: While modern detector arrays can rapidly collect data, \textit{processing} this data is not readily possible at the same rate resulting in large memory requirements and time-consuming post-processing routines. Applications requiring direct or short-term actions (such as in leak detection, chemical process monitoring or medical diagnostics) are then hindered by the time delay between the data acquisition and the processed image.
To overcome this data reduction bottleneck, we show that artificial intelligence (AI) based on a deep convolutional neural network (CNN) \cite{lecun1990, lecun2015, krizhevsky2017} is capable of processing the data generated by the 16'384 hyperspectral pixels in \textit{real-time} on a personal computer, permitting molecule-specific gas live imaging. Built on direct processing of time-domain interferograms, this method extends the AI toolbox for spectroscopy \cite{ji2020, liu2017, acquarelli2016, ghosh2019, gniadecka2004}.

\section{Setup}

\begin{figure}
    \centering
    \includegraphics[width=0.45\textwidth]{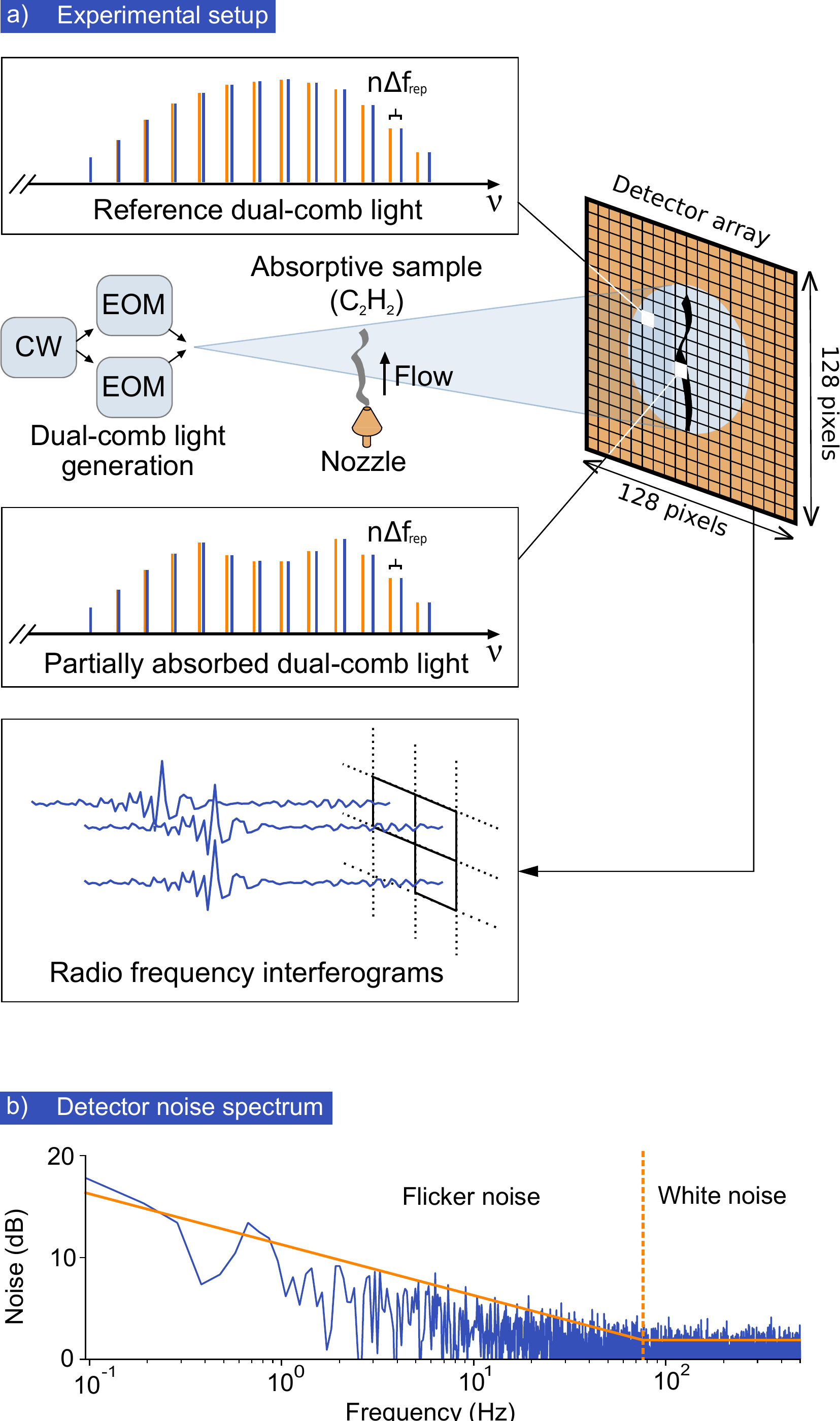}
    \caption{\textbf{Dual-comb hyperspectral imaging and detector noise.}
        \textbf{a.} Highly mutually coherent optical frequency combs are generated by electro-optic modulation (EOM) of a single continuous (CW) wave laser. The dual-comb light is sent through a sample, here a flow of absorbing acetylene (C$_2$H$_2$) gas, then detected by a near- to mid-infrared fast detector array. Each pixel of the $128 \times 128$ detector array simultaneously digitises the dual-comb multi-heterodyne interferograms, which contain spectral information about the sample that can be retrieved via Fourier transformation and normalisation.
        \textbf{b.} Single pixel noise spectrum of the near- to mid-infrared detector array.
    }
    \label{fig1}
\end{figure}

In our experimental setup, a dual-comb source is used to illuminate a near-to-mid-infrared PbSe photo-detector array. The detector array is sensitive to light over the entire 1~to~5~$\mathrm{\upmu}$m wavelength range, which contains the characteristic spectral fingerprint signature of a large number of gas molecules.  
The array, manufactured by NIT, consists of $128 \times 128$~pixels and can be read out with a maximal frame rate of 4~kHz, which corresponds to maximal detectable heterodyne frequency of 2~kHz (Nyquist frequency). Between the dual comb source and the detector, we arrange small nozzles through which acetylene gas (C$_2$H$_2$) can be released, resulting in a jet of gas that is probed by the large diameter dual-comb illumination beam (Figure~\ref{fig1}a). In this way, the spatial structure of the gas flow is projected onto the detector array, where the large number of pixels provides high spatial resolution. 

Analysis of the detector's noise spectrum (Figure~\ref{fig1}b) reveals prohibitively high Flicker noise at low frequencies (below 100~Hz), including in particular pixel dark-value fluctuations. However, following the idea of lock-in detection, the high frame-rate of the detector array provides access to a higher frequency band for heterodyne beatnote detection above 100~Hz that exhibits significantly lower noise. We therefore aim to arrange all heterodyne beatnotes above 100~Hz (and below the Nyquist frequency), thereby overcoming the challenge of low-frequency Flicker noise.

In order to allow dense encoding of spectral information in the desired heterodyne frequency interval, dual-combs of high mutual coherence are required (the width of the heterodyne beatnotes needs to be smaller than their frequency spacing $\Delta f_\mathrm{rep}$). Such high-mutual coherence dual-frequency combs have been demonstrated both in the near- and mid-infrared based on mode-locked lasers, electro-optic modulation and optical parametric oscillators \cite{coddington2008, zolot2012, millot2016, okubo2015, ideguchi2016, mehravar2016, zhao2016, link2017, hebert2018, chen2018, martin-mateos2018, ycas2018,  muraviev2018, kayes2018, liao2018, gu2020, wildi2020, martin-mateos2020, ren2020}. In this work, near-infrared dual-frequency combs are generated via electro-optic modulation\cite{parriaux2020} in a fibre-based setup utilising polarisation maintaining components. Their central wavelength is chosen to coincide with an absorption line of C$_2$H$_2$ gas (spectral line intensity of 4.882$\cdot$10$^{-21}$~cm$\cdot$molecule$^{-1}$) . Specifically, a continuous wave (CW) laser at a wavelength of 1536.7~nm is split into two parts from each of which a comb is derived; this method ensures high-mutual coherence between both combs (Figure~\ref{fig1}a). After splitting, the CW laser's frequency is shifted by 80~MHz and 80~MHz + $\Delta f_\mathrm{c}$ respectively, creating a relative frequency offset $\Delta f_\mathrm{c}$ between the centre frequencies of both combs. Next, optical combs are generated from the shifted CW laser lines by electro-optic modulation (similar to \cite{wildi2020}). In this way, approximately 30 comb lines are generated in each comb and the combs' repetition rates are defined by the electro-optic modulation frequencies. As all modulation sources are referenced to a common 10~MHz clock, high-mutual coherence (heterodyne beatnote width $< 10$~mHz) is readily achieved between both combs. In order to densely sample and resolve the acetylene absorption line (FWHM approx. 10~GHz) we choose $f_\mathrm{rep}^{(1)}=1$~GHz. The detector is operated with a 1~kHz frame rate, which is sufficient in our case and permits recording heterodyne beatnotes up to the Nyquist frequency of 500~Hz. The centre frequency and the spacing of the heterodyne beatnotes are set to $\Delta f_\mathrm{c}=250$~Hz and $\Delta f_\mathrm{rep}=10$~Hz, respectively.

As an aside, we point out that much wider comb spectra can be generated via non-linear spectral broadening \cite{dudley2006}, in particular for a similar configuration of the setup \cite{obrzud2018, obrzud2019}.The detector's maximal frame rate of 4~kHz as well as the sub-Hz linewidth of the heterodyne beatnotes would in principle allow accommodating thousands of spectral channels for a different choice of $\Delta f_\mathrm{c}$ and $\Delta f_\mathrm{rep}$ (a lower/higher value of $\Delta f_\mathrm{rep}$ would entail a reduced/increased acquisition rate).

\begin{figure}
    \centering
    \includegraphics[width=0.45\textwidth]{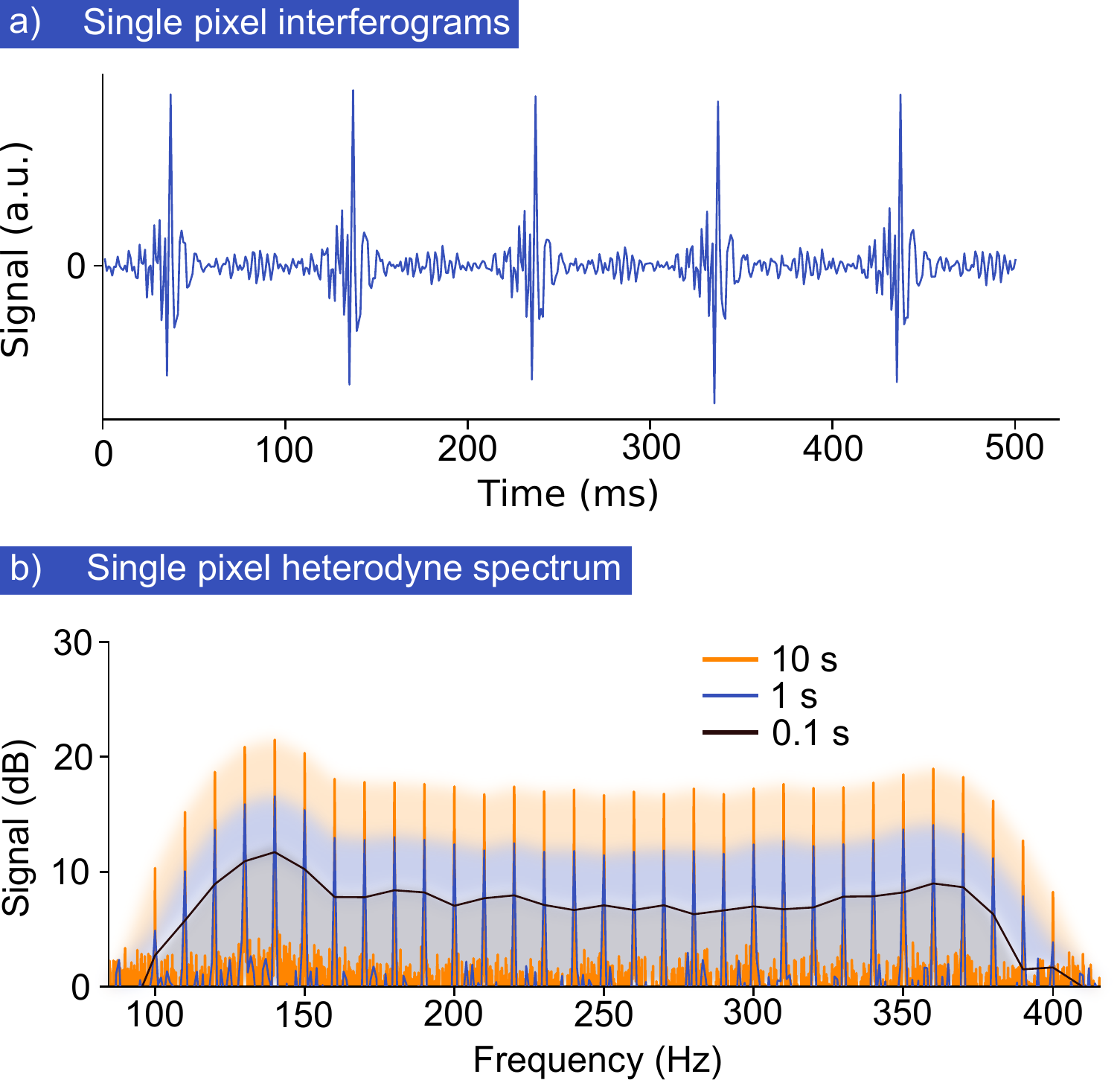}
    \caption{\textbf{Raw interferograms and spectrum.}
        \textbf{a.} Raw dual-comb multi-heterodyne interferograms as recorded by a single pixel.
        \textbf{b.} Multi-heterodyne spectra obtained by Fourier transforming the raw interferograms for different acquisition times (10 s in orange, 1 s in blue, 0.1 s in black).}
    \label{fig2}
\end{figure}

The acquisition rate of 1000 detector frames per second across all 16'384 pixels allows massively parallel acquisition of data, each pixel simultaneously recording the dual-comb time-domain interferograms that contain the spectral information. An example of several interferograms recorded by a \textit{single} pixel of the detector is shown in Figure~\ref{fig2}a (after removal of low frequency components). Fourier-transformation of the raw interferogram trace yields the multi-heterodyne spectrum, as shown in Figure~\ref{fig2}b for different acquisition duration of 0.1~s, 1~s and 10~s (i.e. 1, 10 and 100~interferograms). The spectral envelope of the heterodyne beatnotes reflects the unabsorbed spectral envelope of the dual-combs. For the shortest acquisition time (1~interferogram), the spectral resolution of the heterodyne spectrum corresponds to the frequency spacing of the heterodyne beatnotes. Longer acquisition duration provide higher spectral resolution in the heterodyne spectrum and the heterodyne beatnotes show as narrow spectral peaks. As the higher spectral resolution effectively rejects the incoherent white thermal noise contribution, the signal-to-noise-ratio (SNR) of the heterodyne beatnotes grows proportionally with the square-root of the acquisition duration. %In addition, although not necessary here, the detector array could in principle be cooled for better performance.

Already, single-interferogram acquisition provides a useful SNR of approximately 10. The high-mutual coherence of the combs would in principle permit longer acquisition durations \cite{wildi2020}, and phase-correction \cite{zolot2012, roy2012, ideguchi2014, burghoff2016, hebert2017, zhu2018a,sterczewski2019} can extend this well beyond 1000~s.

\section{Results}
\begin{figure}
    \centering
    \includegraphics[width=0.45\textwidth]{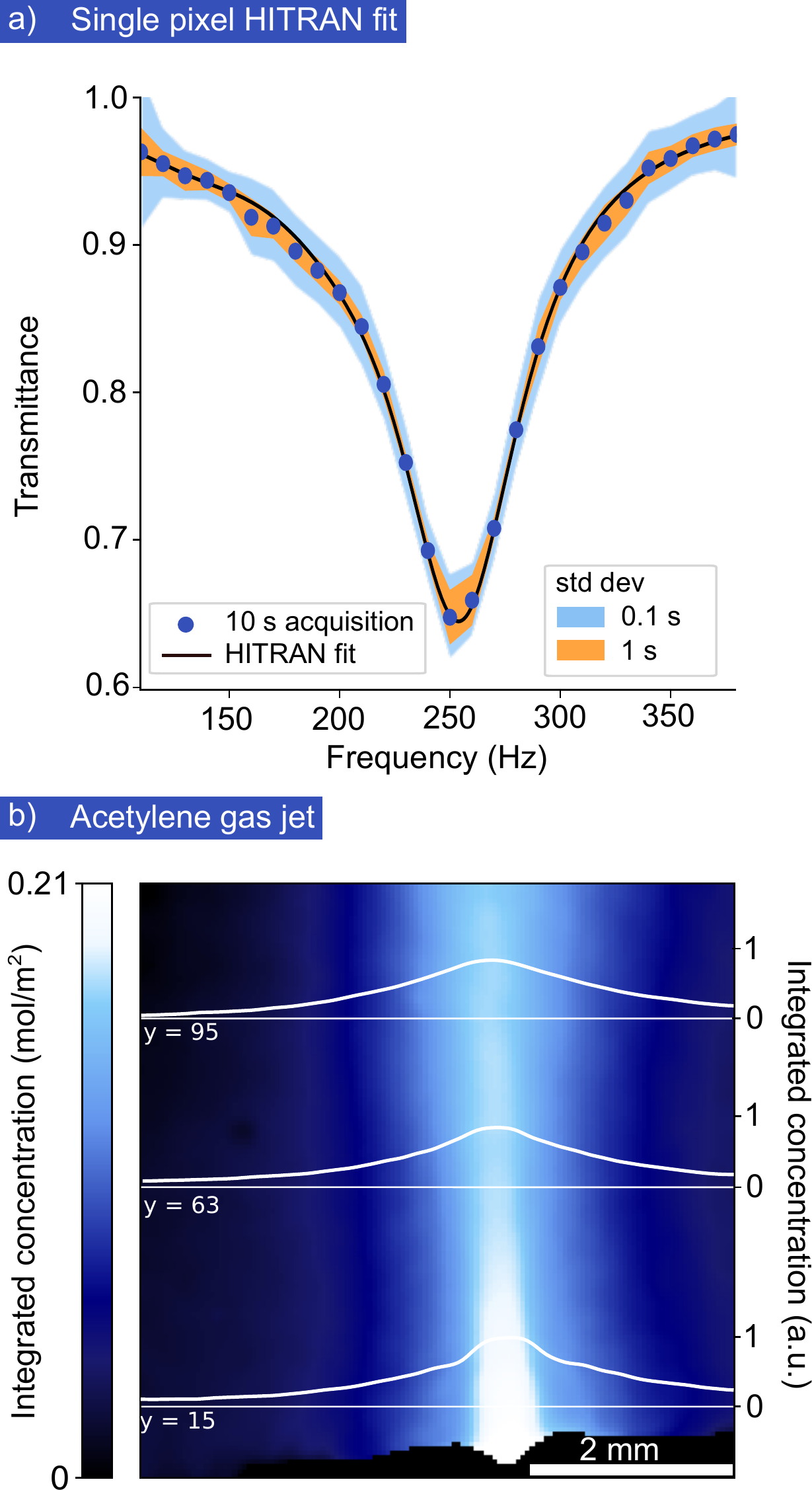}
    \caption{\textbf{Single pixel transmittance spectrum, comparison with HITRAN and integrated concentration image.}
        \textbf{a.} Single pixel dual-comb absorption spectrum of acetylene (C$_2$H$_2$) retrieved from a 10 seconds acquisition (blue dots) compared to a HITRAN fit (black line). The standard deviation of the absorption spectrum for shorter acquisition times (0.1 seconds and 1 second) is shown in bands (centred around the 10 seconds based data).
        \textbf{b.} Reconstructed integrated concentration image of an acetylene flow based on fits to the HITRAN model. Transverse absorption profiles are shown for three different positions along the gas flow (white curves).
    }
    \label{fig3}
\end{figure}

% conventional post-processing
For a first test of hyperspectral imaging, the C$_2$H$_2$ gas jet across the field-of-view is turned on. The heterodyne signal is recorded for each pixel and Fourier-transformed to yield the heterodyne spectra. For each pixel, the transmittance of the dual-comb light on its specific path is obtained by normalising the heterodyne spectrum by an unabsorbed reference heterodyne signal. In our case, the reference heterodyne spectrum is derived from 9 pixels in the top left corner of the detector (where only a negligible amount of gas is present). Alternatively, a pre-recorded reference spectrum may be used. An example of the transmittance signature recorded by a single pixel is shown in (Figure~\ref{fig3}a) for a 10~second long acquisition, showing very good agreement with the HITRAN database (residuals below 3\%). In addition, standard-error bands for shorter acquisition duration (0.1~s and 1~s) are shown, revealing that absorption on the few-percent level can already be detected  based on single-interferogram (0.1~s acquisition duration).

% integrated concentration
To image the spatial distribution of the C$_2$H$_2$ gas, the \textit{integrated concentration} of C$_2$H$_2$ molecules along the light path is derived for each pixel.
Generally, the absorption of light propagating through a sample is described by the Beer-Lambert law
\begin{equation}
    \frac{\mathrm{d}I(\nu, l)}{I(\nu, l)} = - c(l)\cdot m(\nu) \cdot \mathrm{d}l,
\end{equation}
where $I$ is the intensity, $\nu$ is the optical frequency, $l$ is the spatial coordinate along the beam path, $m$ is the molar absorption coefficient and $c(l)$ the molar concentration of gas. Integrating both sides along the path from the dual-comb source to the detector array yields
\begin{equation}\label{absorption_spectrum}
    T(\nu) = \frac{I_d(\nu)}{I_0(\nu)} = \exp\left(-m(\nu) \int_0^L c(l) \mathrm{d}l\right),
\end{equation}
where $T(\nu)$ is the measured transmittance (Figure~\ref{fig3}a), $L$ is the distance between the dual comb source and the detector array, $I_\mathrm{d}(\nu)=I(\nu,L)$ is the intensity at the detector array and $I_0(\nu)=I(\nu, 0)$ is the intensity before absorption.
A natural measure of the number of absorbing particles traversed by the beam is then defined as
\begin{equation}
    C_\mathrm{int} = \int_0^L c(l)\mathrm{d}l
\end{equation}
which can be estimated by fitting the measured transmittance to a HITRAN model.

The integrated concentration $C_\mathrm{int}$ is computed for each pixel based on the 10~second recording (100~interferograms) and shown in (Figure~\ref{fig3}b); the gas jet is well detected and imaged based on its infrared absorption signature. Profiles of $C_\mathrm{int}$ for different heights of the gas flow show a transverse expansion of the gas jet along its flow direction. While in our case only one gas species is imaged, the analysis can readily be generalised to include multiple gas species.

\begin{figure*}
    \centering
    \includegraphics[width=\textwidth]{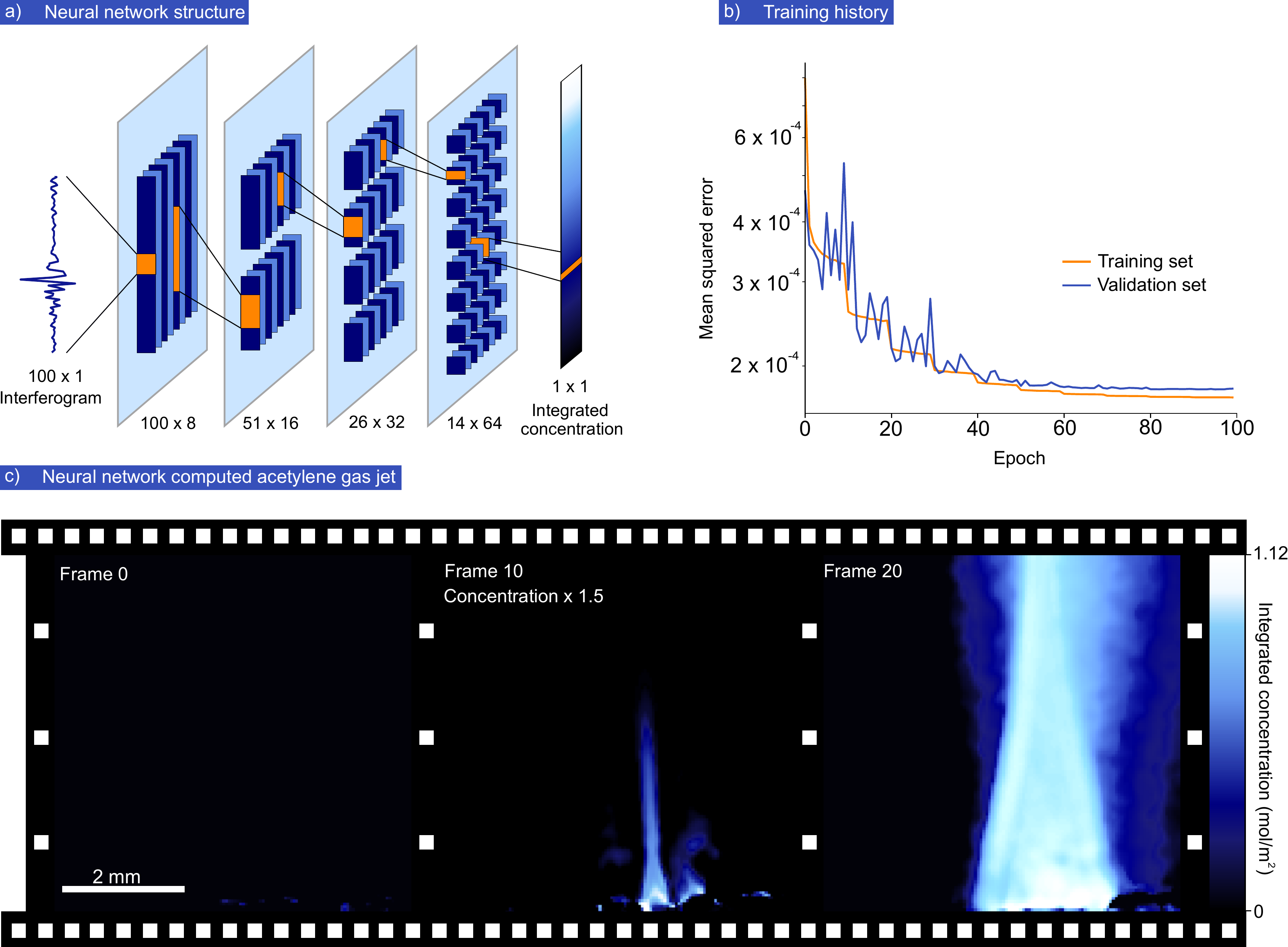}
    \caption{\textbf{Neural network architecture, training history and AI results.}
    \textbf{a.} A single 100 points interferogram is used as input of the convolutional neural network. A first layer with periodic boundary conditions and 8 filters extracts 800 features from the input. Approximately reducing the Kernel width by a factor of two and doubling their number in each subsequent layers keeps the number of features constant across the network. The last layer outputs the integrated concentration corresponding to the input interferogram.
    \textbf{b.} Training history of the network, achieving a standard deviation of 1.3~$\%$ of the maximal integrated concentration with the validation data set.
    \textbf{c.} Three selected frames from a movie that has been reconstructed in real-time by the neural network. The frames show the dynamics of the C$_2$H$_2$ gas jet and are separated by 1~s each. The first frame is recorded before the gas jet is turned on, the second frame depicts the onset of gas emission and the last frame shows the established gas jet.
    }
    \label{fig4}
\end{figure*}

% Transition to AI
Figure~\ref{fig3}b demonstrates that hyperspectral imaging with the near-to-mid-infrared detector array is possible. However, the high data rate of approximately 35~Mb/s and the necessity of performing the analysis on each pixel requires large memory size and result in long computations times, in our case 20 to 30 minutes on a desktop computer for one frame. %Moreover, fitting the transmittance becomes increasingly challenging for shorter acquisition times, where naturally the noise in the retrieved heterodyne spectra increases.

% AI part
Overcoming this data processing bottleneck is crucial in applications demanding fast reactions, such as leak detection or medial diagnostics. To this end, we demonstrate an AI based approach that can directly operate on single (temporal) interferograms and bypasses the conventional analysis presented above. A specifically designed one-dimensional convolutional neural network (CNN) is used for speeding up the processing and implemented using the Keras library running on a Tensorflow backend \cite{abadi2016}. The response of the CNN is invariant against translation of the input data along the time axis; a dedicated trigger or temporal alignment of the input interferogram is not required (any input of a duration of $\Delta f_\mathrm{rep}^{-1}$ is suitable).
This property of CNN greatly simplifies the data analysis and results in better versatility (i.e. applicability to unknown data) of the CNN than densely connected networks with a comparable number of parameters. To avoid any unnecessary processing, the CNN directly takes a single amplitude-normalised interferogram with 100 sampling points as the input (Figure~\ref{fig4}a). The first convolutional layer uses periodic boundary conditions, reflecting the periodic nature of the input and uses a kernel size equal to the size of the input.
In our case, we found that 8~filters on the first layer were sufficient, however, more filters can readily be added to extend the CNNs capabilities to analyse different gas species and mixtures thereof. Each subsequent layer has a kernel size (approximately) divided by two and a doubled number of filters, such that the number of processed features remains constant throughout the layers. Each layer's activation function was chosen to be the rectified linear unit \cite{glorot2011}, as it gave the best performances among a few test functions. The last layer directly provides the scalar integrated concentration value $C_\mathrm{int}$. 
As we detail below, the rapid and massively parallel recording capability of the system allows building large data sets for training such that regularisation layers \cite{srivastava2014} to prevent over-fitting are not necessary. This neural network architecture allows skipping both Fourier transformations as well as time-consuming fitting of absorption profiles.

% Training data set
While the capability of neural networks for fast data processing are widely recognised \cite{lavin2015, cai2016, kozlov2020}, the difficulty of building a reliable \textit{labelled} training data set (containing training input data as well as the correct outcome) is often prohibitive to their use.  In our case, a training data set is rapidly built by sending the comb through a 10~cm gas cell filled with acetylene and arranged between the dual-comb source and the detector (all light traverses the cell). 
Within 30 minutes we record 10 seconds long data sets for 180 different integrated concentrations, ranging from 0 to a maximal integrated concentration of 4.16~mol$\cdot$m$^{-2}$. This way, a set of approximately 300~million interferograms over the full range of values for $C_\mathrm{int}$ is quickly obtained. In order to label each interferogram, i.e. assign to it one of the 180~possible values of $C_\mathrm{int}$, the integrated concentration for each 10 seconds data set is derived as described above via HITRAN fitting. The signal of all pixels is combined for better precision in the derivation of the integrated concentration label. Note that the granularity of the output scale does not degrade the resolution in $C_\mathrm{int}$, which is limited by measurement noise (cf. Figure~\ref{fig3}a). Note that fitting the HITRAN lineshape model is only used for labelling the training data set, not for the actual hyperspectral analysis. Complete independence from a theoretical lineshape model can be achieved if the training data is taken with a known gas concentration.

% Training
Approximately $2\cdot 10^6$ interferograms (equivalent recording time of only 12~s) from the training data set are randomly selected for training and validation of the CNN. The training proceeds over 100 epochs and, on a desktop computer with a standard graphics processing unit, takes approximately 8 hours. A decreasing learning rate divided by 2 every 10 epochs is used to improve the convergence of the learning process. A final standard deviation between the predictions and the expected output of 1.3~$\%$ of the maximum integrated concentration value is reached.

% Performances
To test the CNNs performance, we observe the dynamics of the emerging gas jet when the gas flow is turned on. The multi-heterodyne data of each pixel are processed for each 100~ms time window (single interferogram), so that good temporal resolution is achieved. From the series of reconstructed gas images, three snapshot frames, separated in time by 1~s, are shown in Figure~\ref{fig4}c. In frame 0 no C$_2$H$_2$ gas was released, in frame 10 the gas jet is emerging from one out of several nozzles, and in frame 20 the gas jet from several nozzles is fully developed.
%The small noisy area of non-zero gas concentration values at the bottom of each frame represent areas where no light reached the detector (a situation for which the neural network was not trained).
The results in Figure~\ref{fig4}c show that the CNN can reliably work on single-interferograms (0.1~s acquisitions), permitting the observation of dynamic processes. Importantly, the trained CNN can process the data at a rate that exceeds the raw data recording rate, therefore enabling \textit{real-time} molecule specific imaging with a frame rate of 10~Hz. The CNN also alleviates the need for large memory storage by reducing the heterodyne raw data frame rate from 1000 down to 10 frames per second for the gas images. If desired, the neural network could also be trained to output other parameters e.g. gas temperature (based on line shapes) or be extended to multi-species imaging by adding outputs on the last layer and adjusting the training accordingly.

\section{Conclusion}
In summary, we have shown that dual-comb precision hyperspectral imaging can be performed with an uncooled, high-frame rate near-to-mid-infrared photo-detector array, enabling imaging of gases with molecular specificity. Hyperspectral data has been simultaneously recorded in 16'384 pixels with 30 spectral channels and short acquisition times of 100~ms enabled observation of dynamic phenomena in an acetylene gas jet.
If needed, a significantly larger number of spectral channels could be implemented, at the cost of a reduced image frame rate. 
Key to this demonstration is the high-frame rate of the detector array as well as the high-mutual coherence of the dual-comb illumination, which permits recording the heterodyne signal in a Flicker noise-free frequency band (here above 100~Hz).
Importantly, we have also shown that the high data rate resulting from the massively parallelized hyperspectral data acquisition can be handled in real-time by a convolutional neural network, providing gas concentration images at 10~Hz rate. As the detector array is sensitive across the entire 1~$\upmu$m to 5~$\upmu$m wavelength range, our demonstration can readily be extended to cover the characteristic absorption fingerprints of a wide range of molecular species. Possible extension of our demonstration include the use of high-repetition rate mid-infrared quantum cascade \cite{villares2014, gianella2020} or  microresonator combs\cite{wang2013, yu2018} for broadband spectral imaging of transparent condensed phase media.
\\
\\
{\footnotesize{\textbf{Funding:} This work was supported by the Swiss National Science Foundation (00020\_182598), the Helmholtz Young Investigators Group VH-NG-1404 and the Canton of Neuchâtel.}}

\bibliographystyle{arxiv_mod}
\bibliography{arxiv}

\end{document}